\documentclass{sigchi}

\CopyrightYear{}
\setcopyright{none}
\doi{}
\isbn{}
\conferenceinfo{}{}
\acmPrice{}

\usepackage{balance}       
\usepackage{graphics}      
\usepackage[T1]{fontenc}   
\usepackage{txfonts}
\usepackage{mathptmx}
\usepackage[pdflang={en-US},pdftex]{hyperref}
\usepackage{color}
\usepackage{booktabs}
\usepackage{textcomp}
\usepackage{appendix}
\usepackage{etoolbox}

\usepackage{microtype}        
\usepackage{ccicons}          

\BeforeBeginEnvironment{appendices}{\clearpage}

\def\plaintitle{Quantifying the Chaos Level of Infants' Environment via Unsupervised Learning}
\def\plainkeywords{audio processing, chaos, infant's environments, unsupervised learning, clustering, autoencoders, SOM}
\def\plainauthor{Priyanka Khante, Mai Lee Chang, Domingo Martinez, Kaya de Barbaro, Edison Thomaz}

\makeatletter
\def\url@leostyle{%
  \@ifundefined{selectfont}{
    \def\UrlFont{\sf}
  }{
    \def\UrlFont{\small\bf\ttfamily}
  }}
\makeatother
\def\pprw{8.5in}
\def\pprh{11in}

\definecolor{linkColor}{RGB}{6,125,233}
\hypersetup{
  pdftitle={\plaintitle},
  pdfauthor={\plainauthor},
  pdfkeywords={\plainkeywords},
  pdfdisplaydoctitle=true,
  bookmarksnumbered,
  pdfstartview={FitH},
  colorlinks,
  citecolor=black,
  filecolor=black,
  linkcolor=black,
  urlcolor=linkColor,
  breaklinks=true,
  hypertexnames=false
}

\begin{document}

\title{\plaintitle}

\numberofauthors{3}
\author{%
  \alignauthor{Priyanka Khante\\
    \affaddr{UT Austin}\\
    \affaddr{Austin, Texas, USA}\\
    \email{priyanka.khante@utexas.edu}}\\
  \alignauthor{Mai Lee Chang\\
    \affaddr{UT Austin}\\
    \affaddr{Austin, Texas, USA}\\
    \email{mlchang@utexas.edu}}\\
  \alignauthor{Domingo Martinez\\
    \affaddr{UT Austin}\\
    \affaddr{Austin, Texas, USA}\\
    \email{domingo.martinez@utexas.edu}}\\
  \alignauthor{Kaya de Barbaro\\
    \affaddr{UT Austin}\\
    \affaddr{Austin, Texas, USA}\\
    \email{kaya@austin.utexas.edu}}\\
  \alignauthor{Edison Thomaz\\
    \affaddr{UT Austin}\\
    \affaddr{Austin, Texas, USA}\\
    \email{ethomaz@utexas.edu}}\\
}

\maketitle

\begin{abstract}
  Acoustic environments vary dramatically within the home setting. They can be a source of comfort and tranquility or chaos that can lead to less optimal cognitive development in children. Research to date has only subjectively measured household chaos. In this work, we use three unsupervised machine learning techniques to quantify household chaos in infants' homes. These unsupervised techniques include hierarchical clustering using K-Means, clustering using self-organizing map (SOM), and deep learning. We evaluated these techniques using data from 9 participants which is a total of 197 hours. Results show that these techniques are promising to quantify household chaos.
\end{abstract}

\section{Introduction}
Infants experience a tremendous amount of positive and negative auditory stimulation. Reducing the later and increasing the former can contribute to the healthy mental development of infants and proper acquisition of language \cite{zhao2016effects}. Research evidence suggest that higher levels of chaotic home environment are associated with less optimal cognitive and social development in children \cite{berry2016household}. Research to date has only subjectively measured household chaos \cite{berry2016household}\cite{whitesell2018household}. The ability to be able to predict how chaotic an infant's home is important as preventative measures can be taken in a timely manner to avoid negative consequences. For this project, we propose to use three unsupervised machine learning techniques to classify the intensity of household chaos. These unsupervised techniques include hierarchical clustering using K-Means, clustering using self-organizing map (SOM), and deep learning. Our goal is to predict the intensity of chaos for each 10 second segments of the audio samples. Although, chaos is a very subjective term, we define it as including loud and/or multiple overlapping sounds in the environment. Low chaos refers to distant or soft sounds or sounds that have a tonal quality (e.g. mother talking to the infant, music). We define these terms with respective to the infant and what could be positive or detrimental to the cognitive development of the infant \cite{berry2016household}.

\section{Related Work}
Prior work shows that there is a connection between household chaos and child development \cite{berry2016household}\cite{musacchia2013oscillatory}. Higher levels of chaotic home environment are associated with less optimal cognitive and social development in children \cite{berry2016household}. For example, music and speech patterns with low frequencies have shown to improve the neural development of infants \cite{musacchia2013oscillatory}. Currently, household chaos is only measured qualitatively using surveys \cite{berry2016household}\cite{whitesell2018household}. A common measure decomposed household chaos into household instability and household disorganization \cite{berry2016household}. The other common measure is a survey that included items such as "The children have a regular bedtime routine" and "You can't hear yourself think in our home" \cite{matheny1995bringing}. In order to better understand the effects of household chaos on a child's development, an objective measurement of household chaos is needed to give a deeper insight into the household environments of these households as well as reduce bias from surveys \cite{groves2006nonresponse}\cite{meyer2015household}. 

Within the child development domain, objective measurement methods utilizing machine learning techniques have been explored. For instance, Random Forest have been used to distinguish between cry of preterm and full-term newborns \cite{orlandi2016application}. Support Vector Machines were used to generalize babies of varying ages and vocalization context \cite{aucouturier2011segmentation}. However, machine learning techniques have not been applied to predict the level of chaos in an infant's home environment. In this paper, we show that machine learning techniques are feasible to quantify an infant's household chaos levels.

\section{DATA PRE-PROCESSING AND FEATURE EXTRACTION}
\subsection{Dataset and White Noise Filtering}
We use the dataset from the Baby Wearables Home Study that is provided to us by the Daily Activity Lab supervised by Dr. Kaya De Barbaro. It contains audio samples collected via LENA (Language Environment Analysis) placed on the infant's chest. The data was collected for 26 infants aged six weeks to nine months for a continuous period of 24 hrs each (total of 624 hrs). However, we choose to use a subset of 197 hours for 9 participants (23 hours for 4 and 21 hours for 5) due to computational constraints and missing raw data. For pre-processing, we used version 2.3.0 of Audacity to remove white noise \cite{team2017audacity}. 

\subsection{Audio Feature Extraction}
For each participant, we segment the audio into 1 second audio samples and extract features for each frame. We use a frame size of 1s with 50\% overlap as we want to capture the small nuances in the audio samples. This is a common technique used in many previous environment sound processing works \cite{chu2009environmental}\cite{muhammad2010environment}\cite{cowling2003comparison}. The features extracted are the following and are computed using the Librosa library in Python \cite{mcfee2015librosa}: \textit{raw features} (mean and standard deviation of the raw audio signal), \textit{MFCC}, \textit{root-mean square energy} (RMSE), \textit{zero-crossings}, \textit{spectral centroid}, \textit{spectral bandwidth}, \textit{spectral roll-off} and \textit{spectral flatness}.

To reduce the dimensionality of the features, we compute the mean and standard deviation for the above features for each 1 second frame, giving us a total of 18 features per frame. We further take the average of the means and standard deviations of 20 contiguous segments to get features for 10 second frames with no overlap. This helps us to capture the short bursts of noise while also capturing constant loud noises that are present in the environment. The feature extraction computation took six days on a 12 core computer.

\section{Metric of Evaluation}
\subsection{Silhouette Score}
As we use an unsupervised learning algorithm and we do not have any ground truth regarding the chaos levels of these audio segments, we measure the goodness of our clusters using the silhouette score metric \cite{rousseeuw1987silhouettes}. Silhouette score has a range from -1 to 1 and it measures how similar is a cluster is to itself and how dissimilar is it to other clusters. So a high value closer to 1 signifies a good cluster that is tightly bound and is very different from other clusters. 

\subsection{Random Sampling}
After we get the clusters, we randomly sample segments from each cluster and listen to them and to compute the accuracy of that cluster.

\section{Hierarchical Unsupervised Clustering using K-Means}
\subsubsection{Feature Selection}
For our first model, we use the simple K-means Clustering algorithm \cite{hartigan1979algorithm} with these five chosen features - \textit{raw\_mean} of audio signal, \textit{mfcc\_std} computed on the spectrogram of the audio signal, \textit{rmse\_mean} and \textit{rmse\_std} and \textit{spectral\_centroid\_std}. As the correlation matrix (Figure \ref{fig:correlation_matrix} in Appendix) shows that some of these features are highly correlated with each other, we choose to use a subset of these features which are less than 85\% correlated. To reduce the dimensionality further, we tried out every combination of the features with K-means and decided on these features as they gave the most meaningful results.

\subsubsection{Hierarchical K-Means Clustering}
We tried out various combination of features and used the silhouette score as a metric to judge the goodness of the clustering. However, as it can be seen in Figure \ref{fig:iterateFeatures} (Appendix), for two and three clusters, the silhouette score consistently goes down as you add more features. Therefore, we choose to use the five features, one feature at a time in a hierarchical fashion (\textit{k=2} at each step) to obtain the five clusters as the leaves of the hierarchical tree. The order of the features was determined through trial and error. In this way, we preserve the high silhouette score at branch (only two clusters are produced at each step) and still get meaningful clusters that are able to quantify chaos. This hierarchical tree is depicted in Figure \ref{fig:hierarchicalTree}.

\subsubsection{Results}
\begin{figure}
  \includegraphics[width=0.9\columnwidth]{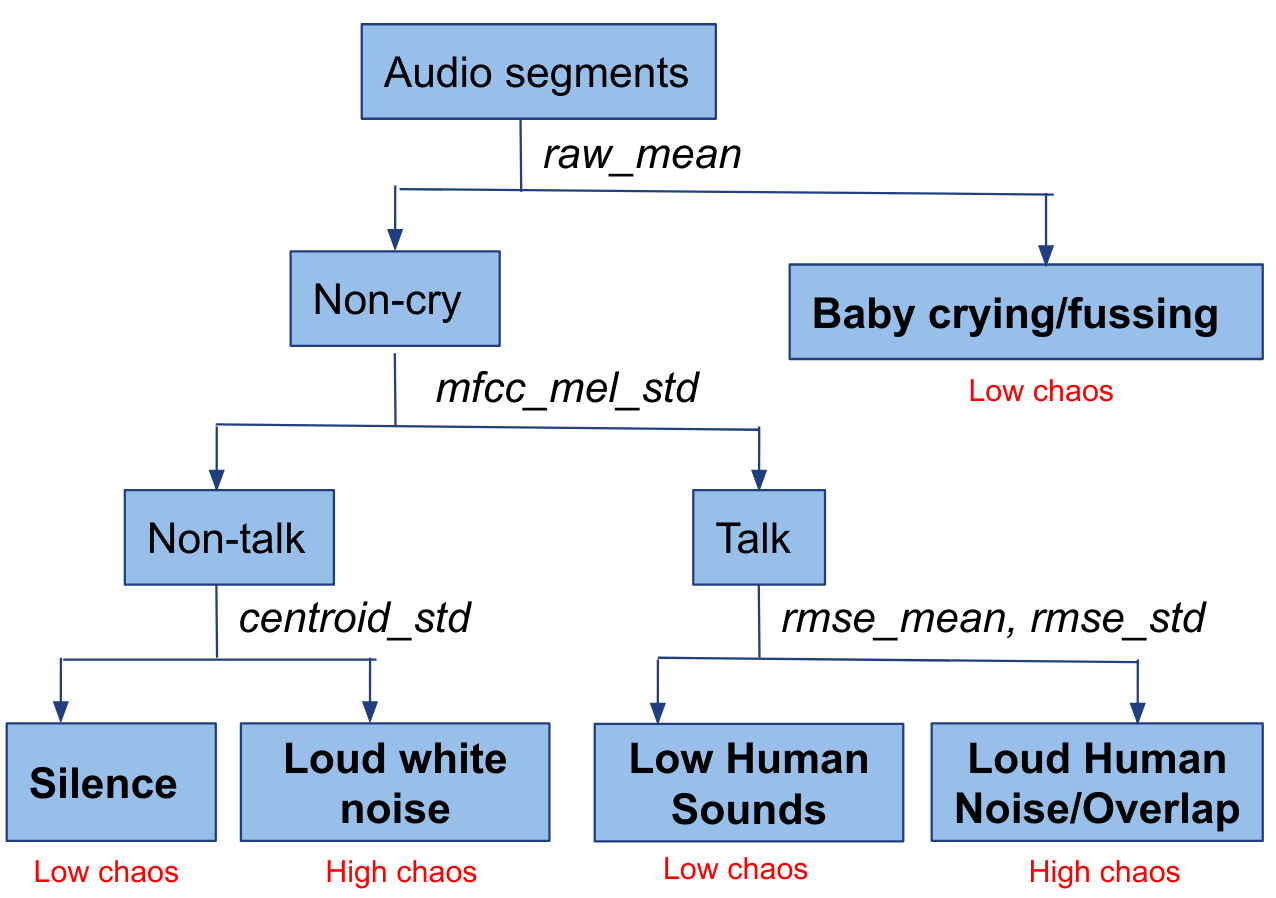}
  \vspace{-3mm}
  \caption{Hierarchical tree obtained by performing KMeans Clustering at each step.}~\label{fig:hierarchicalTree}
  \vspace{-3mm}
\end{figure}

We labelled the five leaf clusters of the tree (Figure \ref{fig:hierarchicalTree}): \textbf{infant crying/fussing}, \textbf{low human sounds} includes conversation with the baby or one person nearby, baby cooing, music, distant TV noise, etc., \textbf{loud human noise/overlap} includes restaurant ambience, parties, loud music, etc. \textbf{silence} includes very low white noise, breathing sounds, sleeping sounds, etc., and \textbf{loud white noise} includes sounds like vacuuming, car passing by, etc.  The silhouette scores obtained at each branch are collated in Table \ref{tab:silhouettehier}.

\begin{table}
  \centering
  \begin{tabular}{c c}
    \hline
    \small\textit{Feature selected}
    & \small \textit{Silhouette Score} \\
    \midrule
    \textit{raw\_mean} & 0.98 \\
    \textit{mfcc\_std} & 0.70 \\
    \textit{rmse\_mean and rmse\_std} & 0.70 \\
    \textit{centroid\_std} & 0.68 \\
  \end{tabular}
  \caption{Silhouette scores obtained at each branch of the Hierarchical Clustering Tree }~\label{tab:silhouettehier}
\end{table}

Figure \ref{fig:hierarchicalTree} also shows the clusters grouped together to show the levels of chaos. As we are classifying chaos with respect to what is detrimental to the infant, we classify \textbf{baby crying/fussing} as low chaos. We classify \textbf{low human sounds} as \textbf{low chaos} as human speech and music have shown to have positive effect on infant's cognitive levels \cite{musacchia2013oscillatory}. We do not care about the misclassified segments during clustering as long as they belong the same level of chaos. For example - baby fussing or cooing sounds classified in the \textbf{low human sounds} cluster are correctly classified as they both are \textit{low chaos} clusters. Figure \ref{fig:bargraph} (Appendix) shows the proportion of time spent by 4 of the households in each cluster. It can be seen that the time spent in each chaos levels varies and it would interesting future work to see how this affects the infant's cognitive skills.

In addition to the silhouette score, we also sampled 50 audio segments from each of these clusters and computed the accuracy of the cluster.We calculate the accuracy of our clusters, by counting the number of correctly classified segments and dividing by 50, and these results are tabulated in Table \ref{tab:clusteraccuracy}. Taking an average over these accuracy, this model has a 81.2\% accuracy. The \textbf{loud human noise/overlap} cluster has the lowest accuracy as most of the misclassified segments were that of the baby crying. Acoustically, it makes sense that they fall in the same cluster as they are equally loud and chaotic. However, in this context, baby crying should be \textit{low chaos} and hence leads to misclassification. In future work, we plan to remove baby noises before clustering to improve our results.

\begin{table}
  \centering
  \begin{tabular}{c c}
    \hline
    {\small\textit{Cluster}}
    & {\small \textit{Cluster Accuracy}} \\
    \midrule
    \textit{Baby Crying/Fussing} & 0.90 \\
    \textit{Low Human Sounds} & 0.94 \\
    \textit{Loud Human Noise/Overlap} & 0.56 \\
    \textit{Silence} & 0.82 \\
    \textit{Loud White Noise} & 0.84 \\
  \end{tabular}
  \caption{Cluster Accuracy with Random Sampling of 50 segments from each cluster}~\label{tab:clusteraccuracy}
\end{table}

\section{Unsupervised Clustering using Self-Organizing Map}
The self-organizing map (SOM) is a type of neural network that transforms the input space into a two-dimensional map. The input space are the training samples fed into the neural network, and the neurons connected to the training samples represent the nodes of the two-dimensional grid. Given that the layer of neurons are the nodes in the two-dimensional map, we will use neurons and nodes interchangeably. As a result, a map is formed which reduces the dimensionality of the input space (dataset). The foundation of the SOM resides in the classical vector quantization (VQ) which creates "optimally tuned feature-sensitive filters by competitive learning" \cite{kohonen2013essentials}. However, Kohonen et al. \cite{kohonen1982self}, adjusted the VQ models to become spatially and globally ordered. Doing so allows similar models to be closer on the grid, whereas models that are dissimilar are farther away on the grid. Therefore, as input vectors(batches) are continuously fed into the neural network, the position of the nodes is adjusted on the map as the weights for the input vector are adjusted over time. 

In SOM, there are no activation functions and the weights are a characteristic of the node itself. The weight can be considered as a coordinate in the input space. The columns of the input space are the number of weight coordinates for each node. For instance, if there were 4 columns in the input space, then each node would have its own 4-dimensional coordinate point. 

\subsubsection{Feature Selection}
In this research, we elected to utilize 18 features to create an 18x18 grid that produces 324 neurons (or nodes). After generating the SOM, we can see the weight planes for each feature for the entire data set. In Figure \ref{fig:figure1} (Appendix), lighter colors represent smaller weights and weight planes that are similar indicate features that are highly correlated. For instance, input 3 (Flatness Mean) and input 4 (Flatness Standard Deviation) are highly correlated. Based on the information from Figure \ref{fig:figure1} (Appendix), we can eliminate several redundant features to reduce the dimensionality of our model and increase its performance. In this model, 18 features have been reduced to 8 and the number of nodes has been reduced from 324 to 256, and again to 16. Figure \ref{fig:figure2} (Appendix) shows similar patterns for 256 Neurons as well as 16 Neurons, but the number of nodes has been drastically reduced so that a reasonable number of clusters can be generated. Therefore, this model will generate 16 varying levels of chaotic a home environment.

\subsubsection{Weighted Planes}
Looking at neighbor weight distances in Figure \ref{fig:figure3} (Appendix), we can visualize the Euclidean distance for each node with respect to its neighbor. Darker connections represent features in the input space that are far apart while lighter colors represent nodes in close proximity. Therefore, a series of dark connections can be considered as barriers that separate larger regions of similar features. However, in our model there are few dark borders but several regions where red and orange overlap to separate themselves from yellow. This represents a close proximity between nodes corresponding to features that share similar characteristics. This suggest varying overlap in the type of sound classes experienced for all participants. In Figure \ref{tab:classifieraccuracy}, the 256 Neuron case shows two runs for the same model but seemingly different results. However, taking a closer look we can see the patterns generated are similar but only the orientation is different, and the same patterns are generated for the 16-Neuron case.

\subsubsection{Clustering}
Clustering of sound classification occurs on multiple dimensions of the input space. For instance, Figure \ref{fig:figure20}(Appendix) shows two 2-dimensional clustering which provides information for 3 of the 8 dimensions in the input space. The unsupervised model generated 16 clusters which were inputted into a supervised learning classifier to verify the accuracy of each cluster. We elected to use 5-fold cross validation to prevent over-fitting rather than leave one out since we are inputting data from a shallow learning architecture. We compared both Weighted K-Nearest Neighbors and Fine Decision Trees to determine the highest accuracy. Additionally, to help further reduce the dimensionality of the dataset, we also compared both classifiers using principal component analysis with 95 percent variance. Table \ref{tab:classifieraccuracy} shows the accuracy for each classifier. 
\subsubsection{Limitations and Further Work}
Although clusters were generated and its accuracy provided, not all clusters were verified to indicate very similar sounds. However, the clusters that were verified provided similar sounds so clusters must be labeled, and those labeled clusters must be organized chaos levels. 

\begin{table}
  \centering
  \begin{tabular}{c c}
    \hline
    {\small\textit{Classifier}}
    & {\small \textit{Accuracy}} \\
    \midrule
    \textit{Decision Tree} & 0.88 \\
    \textit{KNN} & 0.97 \\
    \textit{Decision Tree (with PCA)} & 0.92 \\
    \textit{KNN (with PCA)} & 0.96 \\
  \end{tabular}
  \caption{Classifier Accuracy}~\label{tab:classifieraccuracy}
\end{table}

\section{Unsupervised Deep Learning}
\begin{figure}
  \includegraphics[width=0.9\columnwidth]{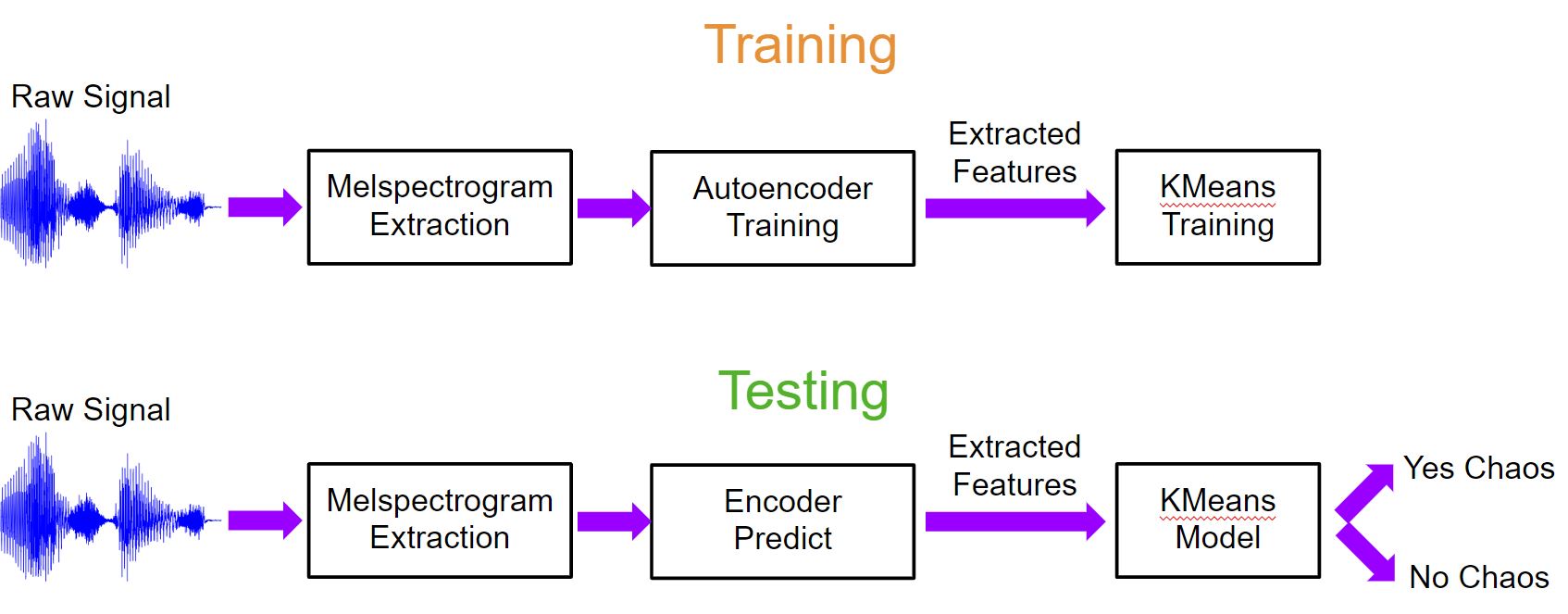}
  \vspace{-3mm}
  \caption{Training and testing process for the unsupervised deep learning approach}~\label{fig:dl_process}
  \vspace{-3mm}
\end{figure}

We also explored using an unsupervised deep learning approach to distinguish audio signals that are chaotic or not (termed "yes chaos" and "no chaos"). We used Keras with TensorFlow as the backend. Figure \ref{fig:dl_process} shows the training and testing process for this approach. During the training phase, we first normalized the raw data and then extracted the melspectrogram. We also normalized the melspectrogram and fed this into the autoencoder to automatically extract features. We trained the autoencoder and used the extracted features as inputs to the KMeans model. In the testing phase, the difference is that we use the encoder to predict the features and use these features to the trained KMeans model to distinguish between yes and no chaos. 

\subsection{Autoencoder Design}
\begin{figure}
  \includegraphics[width=0.9\columnwidth]{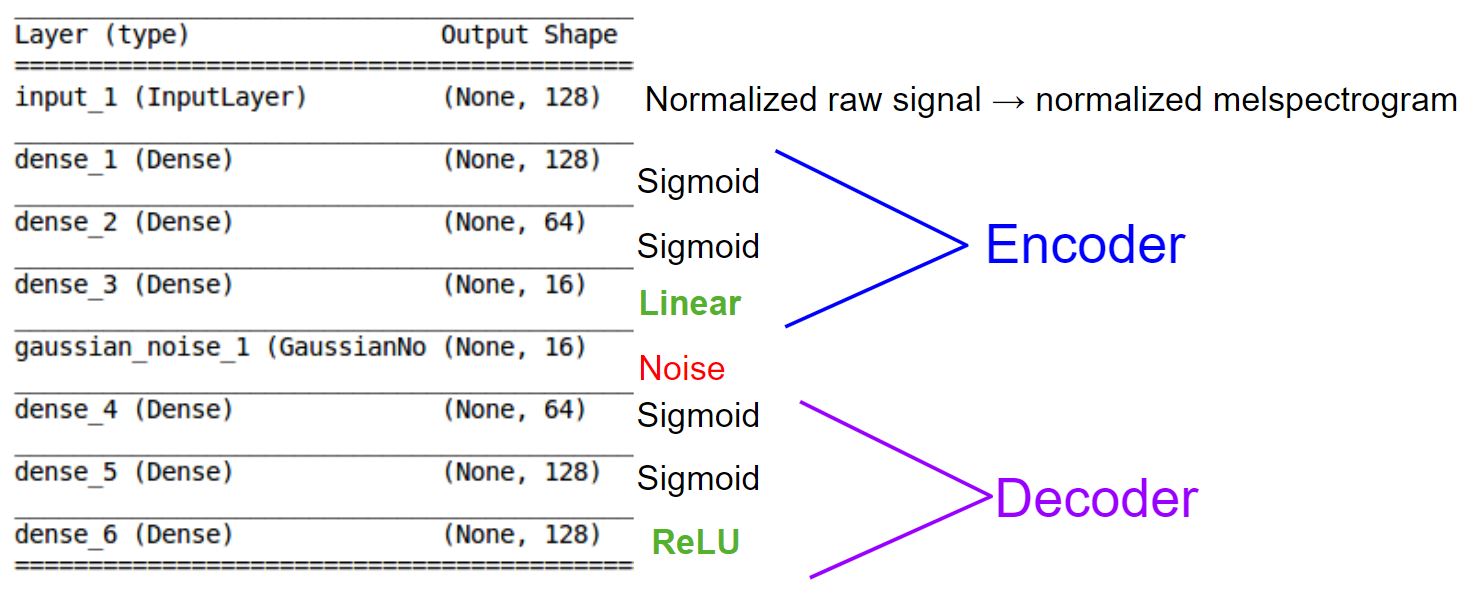}
  \vspace{-3mm}
  \caption{Design of the autoencoder}~\label{fig:auto_design}
  \vspace{-3mm}
\end{figure}

Autoencoders are used for automatic feature extraction. It is trained to recreate the original input by using a smaller representation. The autoencoder includes two main components, the encoder and decoder. The encoder coverts the input to a lower dimensional representation and the the decoder uses this representation to recreate the original input. We designed a fully connected, symmetric autocoder with three layers each for the encoder and decoder (Figure \ref{fig:auto_design}). All the layers are a feed-forward neural network. For the encoder, the first and second layers, the activation function is sigmoid. The activation function for the third layer is linear. The rationale for choosing linear is that we want the features to span a dynamic range as large as possible. Next, we added Gaussian noise ($\mu = 0, \sigma = 1/\sqrt{10}$) in order to regularize the model and train the encoder to generate separable features which helps the clustering model later. For the encoder layers, the activation functions are sigmoid, sigmoid, and ReLu. The rationale for choosing ReLu is that it matches the positive values of the input data. 

\subsubsection{Parameters Selection}
To select the autoencoder parameters, we trained 9 models. We varied the latent dimensions (8, 16, 32) and the adadelta optimizer learning rate (0.1, 1, 10). We used mean squared error for the loss, window size of 1 second (8,000 samples) with no overlap, batch size of 4096, 10,000 epochs, and 90\% train and 10\% validation data. The input data is taken from two participants where it is balanced in terms of silence and non-silence (5 hrs of silence and 5 hrs of non-silence from each participant). Examples of non-silence are people talking and music/TV in the background. Examples of silence include white noise and static noise. Figure \ref{fig:auto_params} (Appendix) shows the results where the best latent dimension is 16 and learning rate is 10 (training loss = 0.03).

\begin{figure}
  \includegraphics[width=0.9\columnwidth]{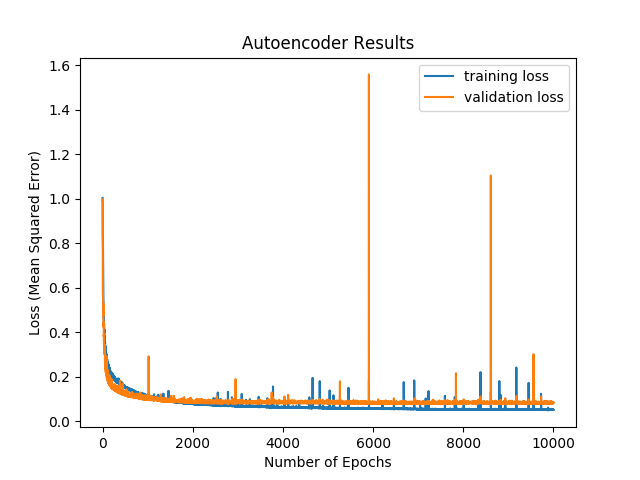}
  \caption{Training result of autoencoder using adadelta as the optimizer (learning rate =10)}~\label{fig:adadelta}
\end{figure}

\begin{figure}
  \includegraphics[width=0.9\columnwidth]{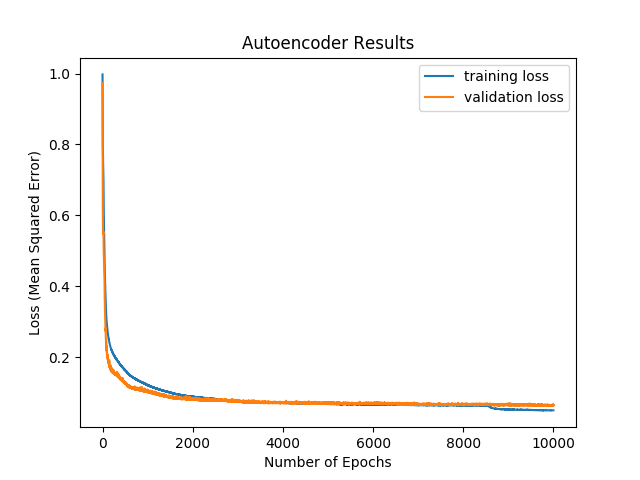}
  \vspace{-3mm}
  \caption{Training result of autoencoder using RMSprop as the optimizer}~\label{fig:rms}
  \vspace{-3mm}
\end{figure}

Based on the results of exploring the parameters, we trained the autoencoder with balanced data from four participants which is a total of 40 hrs. We used 90\% train and 10\% test. We also used mean squared error for the loss, window size of 1 sec with no overlap, batch size of 4096, 10,000 epochs, 16 latent dimensions, and adadelta with a learning rate of 10. Results show that the training loss is 0.05 and validation loss is 0.08. However, the results plot (Figure \ref{fig:adadelta}) show that the validation loss spiked a few times. Closer investigation revealed that the audio segments with the spikes are when there is close to complete silence, thus adadelta is not good at learning this type of data. Next we modified the model to use a different optimizer, RMSprop. Results in Figure \ref{fig:rms} showed that RMSprop can better handle the close to complete silence audio (training loss = 0.05, validation loss = 0.06)

\subsection{KMeans Model Training}
Next, we used the extracted features as input to train the KMeans model. The number of clusters is unknown so we experimented with various sizes (2, 3, 4, 6, 8, 12, 16) and calculated the silhouette score. Results in Table \ref{tab:table_kmeans} show that the silhouette scores are similar to each other. Step plots of the results (Figures \ref{fig:2clusters} and \ref{fig:12clusters} in Appendix) show that as the number of clusters increases, i.e., more granularity, the model is able to better distinguish between silence and non-silence.  

\begin{table}
  \centering
  \begin{tabular}{c c}
    \hline
    {\small\textit{Number of Clusters}}
    & {\small \textit{Silhouette Score}} \\
    \midrule
    2 & 0.89 \\
    3 & 0.88 \\
    4 & 0.88 \\
    6 & 0.86 \\
    8 & 0.83 \\
    12 & 0.78 \\
  \end{tabular}
  \caption{KMeans model training results.}~\label{tab:table_kmeans}
\end{table}

In practice, a system that clusters every 1 second may be at a scale that is too fine. Thus, we created a 11 second window with majority vote. Within the 11 second window, there will be a yes/no chaos prediction for each second. The prediction that is the most frequent is chosen as the final prediction for the entire window.  

\subsection{Results}
For testing, we used balanced data that has not been seen by the encoder. This data is taken from four participants (66 seconds per participant). To assess the system's accuracy, we manually listened to the audio files. For the KMeans model with 2 clusters, the accuracy is about 92\%. For the KMeans model with 3 clusters, the accuracy is about 96\%. With the 3 clusters, one prediction is no chaos and the other two predictions are yes chaos.  

\section{Contribution}
Research to date has only subjectively measured household chaos \cite{berry2016household}\cite{whitesell2018household}. Our work contributes by showcasing that unsupervised learning approaches can be used to classify levels of chaos. To our knowledge, this is a first work in the literature that tries to qualitatively quantify household chaos. We explored unsupervised techniques include hierarchical clustering using K-Means, clustering using SOM, and deep learning and show that these techniques are promising in objectively quantifying household chaos to reduce the burden on hand annotations as well as bias from surveys. The three different models were all successful in classifying chaos albeit to different levels of granularity. This gives the user the power to choose the model depending on level of granularity of required. 

\section{Discussion and Future Work}
We show that unsupervised hierarchical clustering using K-Means, clustering using SOM, and deep learning are able to successfully classify between chaos and non-chaos. However, as the work is unsupervised it is hard to compute the accuracy of clusters formed and has to be subjected to individual's opinion. Future work will involve calculating inter-rater reliability and implementing semi-supervised techniques, e.g. active learning, to improve our results. In addition, manual feature extraction took about six days in which future work includes ways to reduce this computation time. With the deep learning approach, a common challenge is the interpretability of the extracted features. Deciding which approach to use will depend on how the system is used, i.e., importance of false alarms.  

The other limitation of our work is that the segments include baby crying and other baby sounds, which should be removed as we are trying to classify the environmental sounds. In future work, we plan to remove these segments as they reduce the accuracy of our models and retrain our models. 

Also, as the data collected is from participants with a high socioeconomic income background, it is not very representative of all households. Therefore, more data needs to be collected in order to train better models and get a more balanced data in order to quantitatively conclude about the effect of chaos on infants.

\section{Conclusion}
In this work, we show that unsupervised learning approaches (hierarchical clustering using K-Means, clustering using SOM, and deep learning) can be used to qualitatively classify the different intensities of chaos in household environments thus reducing the burden on hand annotations and bias in surveys. We designed, trained, validated, and tested the three approaches on data from 9 participants which were at least successfully able to differentiate between chaos and non-chaos for 10 second audio segments. Our results show that this approach is promising and future work will use these classified intensities of chaos to look into cognition development of infants.

\pagebreak{}

\bibliographystyle{SIGCHI-Reference-Format}
\bibliography{sample}


\begin{thebibliography}{00}


\ifx \showCODEN    \undefined \def \showCODEN     #1{\unskip}     \fi
\ifx \showDOI      \undefined \def \showDOI       #1{{\tt DOI:}\penalty0{#1}\ }
  \fi
\ifx \showISBNx    \undefined \def \showISBNx     #1{\unskip}     \fi
\ifx \showISBNxiii \undefined \def \showISBNxiii  #1{\unskip}     \fi
\ifx \showISSN     \undefined \def \showISSN      #1{\unskip}     \fi
\ifx \showLCCN     \undefined \def \showLCCN      #1{\unskip}     \fi
\ifx \shownote     \undefined \def \shownote      #1{#1}          \fi
\ifx \showarticletitle \undefined \def \showarticletitle #1{#1}   \fi
\ifx \showURL      \undefined \def \showURL       #1{#1}          \fi

\bibitem{aucouturier2011segmentation}
{Jean-Julien Aucouturier}, {Yulri Nonaka}, {Kentaro Katahira}, {and} {Kazuo
  Okanoya}. 2011.
\newblock \showarticletitle{Segmentation of expiratory and inspiratory sounds
  in baby cry audio recordings using hidden Markov models}.
\newblock {\em The Journal of the Acoustical Society of America\/} {130}, 5
  (2011), 2969--2977.
\newblock


\bibitem{berry2016household}
{Daniel Berry}, {Clancy Blair}, {Michael Willoughby}, {Patricia
  Garrett-Peters}, {Lynne Vernon-Feagans}, {W~Roger Mills-Koonce}, {Family Life
  Project~Key Investigators}, {and} {others}. 2016.
\newblock \showarticletitle{Household chaos and children's cognitive and
  socio-emotional development in early childhood: Does childcare play a
  buffering role?}
\newblock {\em Early childhood research quarterly\/}  {34} (2016), 115--127.
\newblock


\bibitem{chu2009environmental}
{Selina Chu}, {Shrikanth Narayanan}, {and} {C-C~Jay Kuo}. 2009.
\newblock \showarticletitle{Environmental sound recognition with
  time--frequency audio features}.
\newblock {\em IEEE Transactions on Audio, Speech, and Language Processing\/}
  {17}, 6 (2009), 1142--1158.
\newblock


\bibitem{cowling2003comparison}
{Michael Cowling} {and} {Renate Sitte}. 2003.
\newblock \showarticletitle{Comparison of techniques for environmental sound
  recognition}.
\newblock {\em Pattern recognition letters\/} {24}, 15 (2003), 2895--2907.
\newblock


\bibitem{groves2006nonresponse}
{Robert~M Groves}. 2006.
\newblock \showarticletitle{Nonresponse rates and nonresponse bias in household
  surveys}.
\newblock {\em Public opinion quarterly\/} {70}, 5 (2006), 646--675.
\newblock


\bibitem{hartigan1979algorithm}
{John~A Hartigan} {and} {Manchek~A Wong}. 1979.
\newblock \showarticletitle{Algorithm AS 136: A k-means clustering algorithm}.
\newblock {\em Journal of the Royal Statistical Society. Series C (Applied
  Statistics)\/} {28}, 1 (1979), 100--108.
\newblock


\bibitem{kohonen1982self}
{Teuvo Kohonen}. 1982.
\newblock \showarticletitle{Self-organized formation of topologically correct
  feature maps}.
\newblock {\em Biological cybernetics\/} {43}, 1 (1982), 59--69.
\newblock


\bibitem{kohonen2013essentials}
{Teuvo Kohonen}. 2013.
\newblock \showarticletitle{Essentials of the self-organizing map}.
\newblock {\em Neural networks\/}  {37} (2013), 52--65.
\newblock


\bibitem{matheny1995bringing}
{Adam~P Matheny~Jr}, {Theodore~D Wachs}, {Jennifer~L Ludwig}, {and} {Kay
  Phillips}. 1995.
\newblock \showarticletitle{Bringing order out of chaos: Psychometric
  characteristics of the confusion, hubbub, and order scale}.
\newblock {\em Journal of Applied Developmental Psychology\/} {16}, 3 (1995),
  429--444.
\newblock


\bibitem{mcfee2015librosa}
{Brian McFee}, {Colin Raffel}, {Dawen Liang}, {Daniel~PW Ellis}, {Matt
  McVicar}, {Eric Battenberg}, {and} {Oriol Nieto}. 2015.
\newblock \showarticletitle{librosa: Audio and music signal analysis in
  python}. In {\em Proceedings of the 14th python in science conference}.
  18--25.
\newblock


\bibitem{meyer2015household}
{Bruce~D Meyer}, {Wallace~KC Mok}, {and} {James~X Sullivan}. 2015.
\newblock \showarticletitle{Household surveys in crisis}.
\newblock {\em Journal of Economic Perspectives\/} {29}, 4 (2015), 199--226.
\newblock


\bibitem{muhammad2010environment}
{Ghulam Muhammad}, {Yousef~A Alotaibi}, {Mansour Alsulaiman}, {and}
  {Mohammad~Nurul Huda}. 2010.
\newblock \showarticletitle{Environment recognition using selected MPEG-7 audio
  features and Mel-Frequency Cepstral Coefficients}. In {\em Digital
  Telecommunications (ICDT), 2010 Fifth International Conference on}. IEEE,
  11--16.
\newblock


\bibitem{musacchia2013oscillatory}
{Gabriella Musacchia}, {Naseem~A Choudhury}, {Silvia Ortiz-Mantilla}, {Teresa
  Realpe-Bonilla}, {Cynthia~P Roesler}, {and} {April~A Benasich}. 2013.
\newblock \showarticletitle{Oscillatory support for rapid frequency change
  processing in infants}.
\newblock {\em Neuropsychologia\/} {51}, 13 (2013), 2812--2824.
\newblock


\bibitem{orlandi2016application}
{Silvia Orlandi}, {Carlos Alberto~Reyes Garcia}, {Andrea Bandini}, {Gianpaolo
  Donzelli}, {and} {Claudia Manfredi}. 2016.
\newblock \showarticletitle{Application of pattern recognition techniques to
  the classification of full-term and preterm infant cry}.
\newblock {\em Journal of Voice\/} {30}, 6 (2016), 656--663.
\newblock


\bibitem{rousseeuw1987silhouettes}
{Peter~J Rousseeuw}. 1987.
\newblock \showarticletitle{Silhouettes: a graphical aid to the interpretation
  and validation of cluster analysis}.
\newblock {\em Journal of computational and applied mathematics\/}  {20}
  (1987), 53--65.
\newblock


\bibitem{team2017audacity}
{Audacity Team}. 2017.
\newblock Audacity: Free Audio Editor and Recorder (Version 2.1. 3)[Computer
  software].
\newblock   (2017).
\newblock


\bibitem{whitesell2018household}
{Corey~J Whitesell}, {Brian Crosby}, {Thomas~F Anders}, {and} {Douglas~M Teti}.
  2018.
\newblock \showarticletitle{Household chaos and family sleep during infants'
  first year.}
\newblock {\em Journal of Family Psychology\/} {32}, 5 (2018), 622.
\newblock


\bibitem{zhao2016effects}
{T~Christina Zhao} {and} {Patricia~K Kuhl}. 2016.
\newblock \showarticletitle{Effects of enriched auditory experience on infants'
  speech perception during the first year of life}.
\newblock {\em Prospects\/} {46}, 2 (2016), 235--247.
\newblock


\end{thebibliography}

\begin{appendices}
\section{APPENDIX}   
\begin{figure}[!htb]
  \includegraphics[width=1.1\columnwidth]{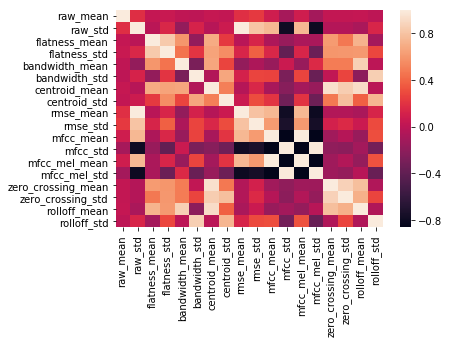}
  \vspace{-3mm}
  \caption{Correlation Matrix for the Audio Features}~\label{fig:correlation_matrix}
  \vspace{-3mm}
\end{figure}

\begin{figure}[!htb]
  \includegraphics[width=1.0\columnwidth]{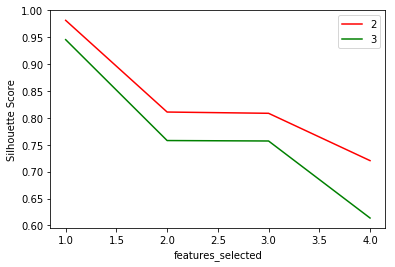}
  \caption{Plot of features selected against the silhouette score obtained. The score drops for each features that gets added. }~\label{fig:iterateFeatures}
\end{figure}

\begin{figure}[!htb]
  \includegraphics[width=1.0\columnwidth]{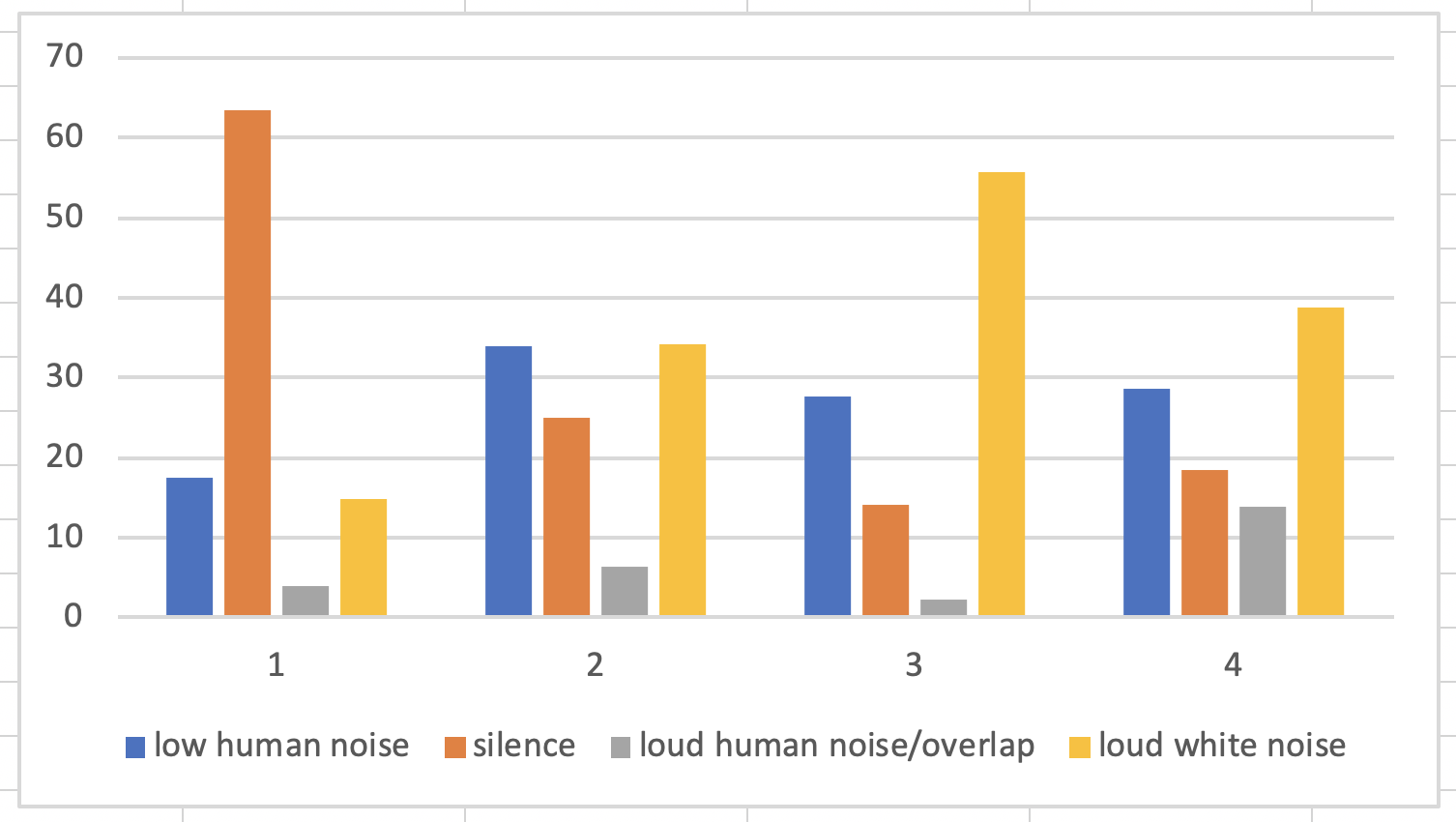}
  \vspace{-3mm}
  \caption{Proportion of time spent in each cluster for 4 participant households. We do not show \textbf{baby crying/fusssing} as we care about the infant's environment and not the sounds made by the infant itself. }~\label{fig:bargraph}
  \vspace{-3mm}
\end{figure}

\begin{figure}[!htb]
\centering
 \includegraphics[width=0.9\columnwidth]{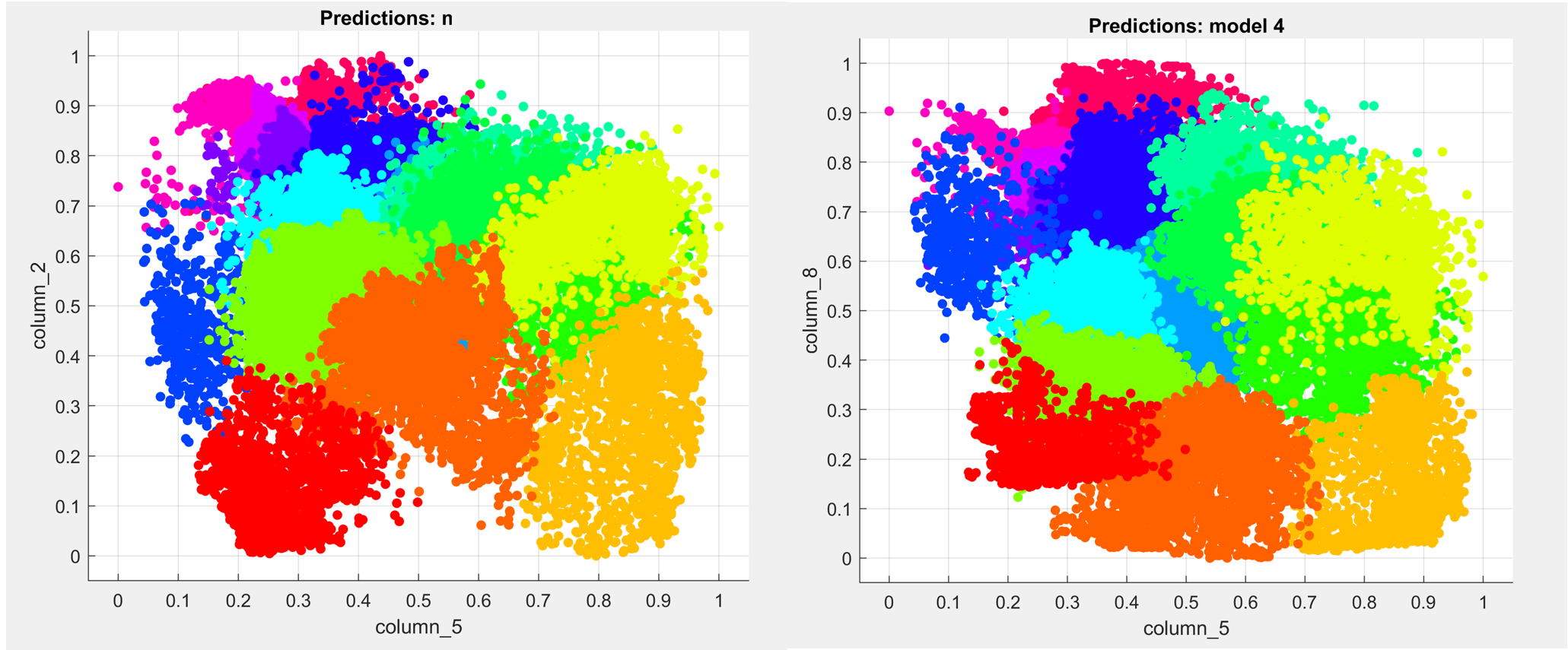}
  \caption{16 Neuron Clusters}~\label{fig:figure20}
\end{figure}
\begin{figure}[!htb]
  \includegraphics[width=0.9\columnwidth]{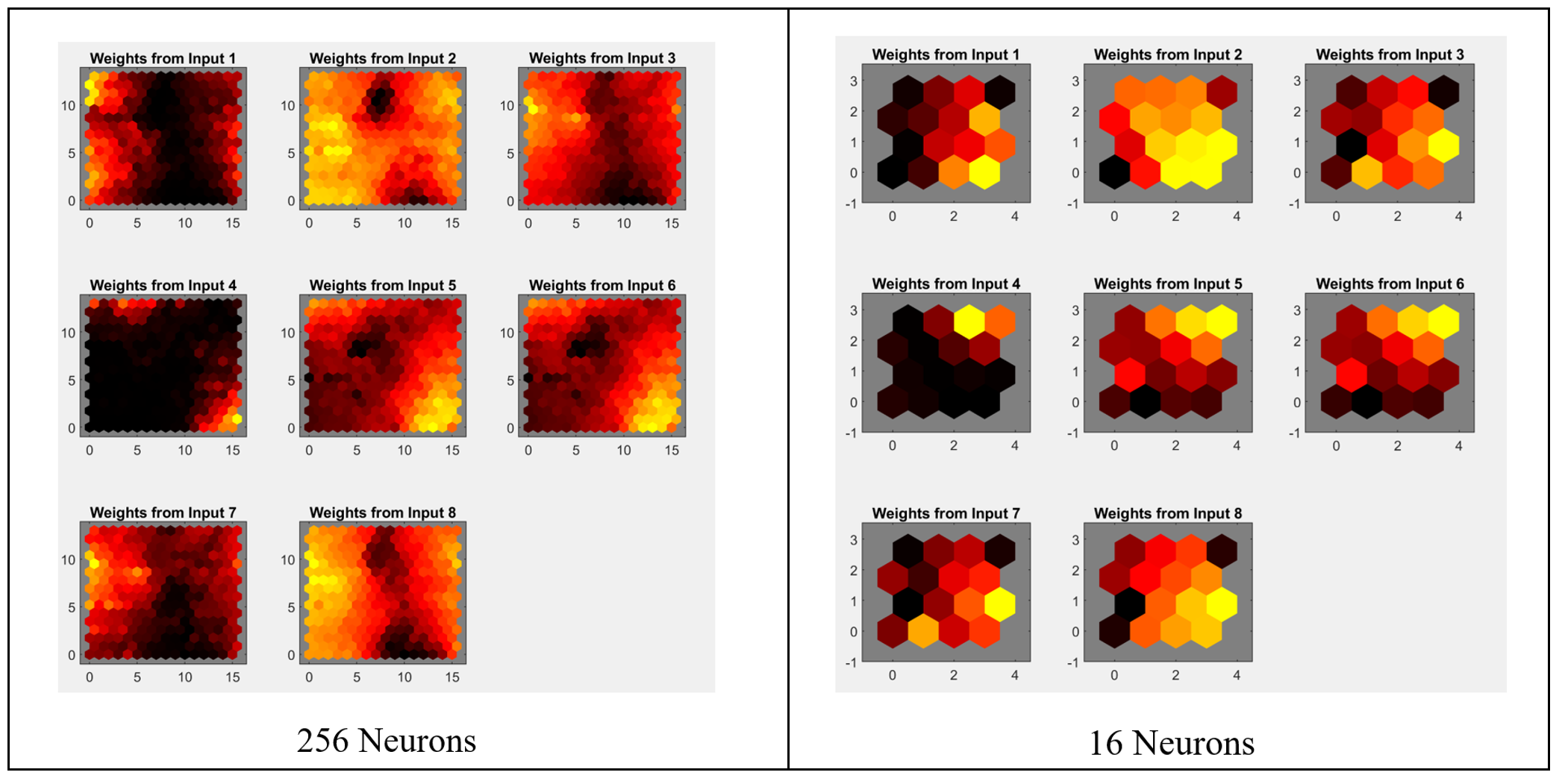}
  \vspace{-3mm}
  \caption{Weight Planes for reduced input space}~\label{fig:figure2}
  \vspace{-3mm}
\end{figure}
\begin{figure}[!htb]
  \includegraphics[width=0.9\columnwidth]{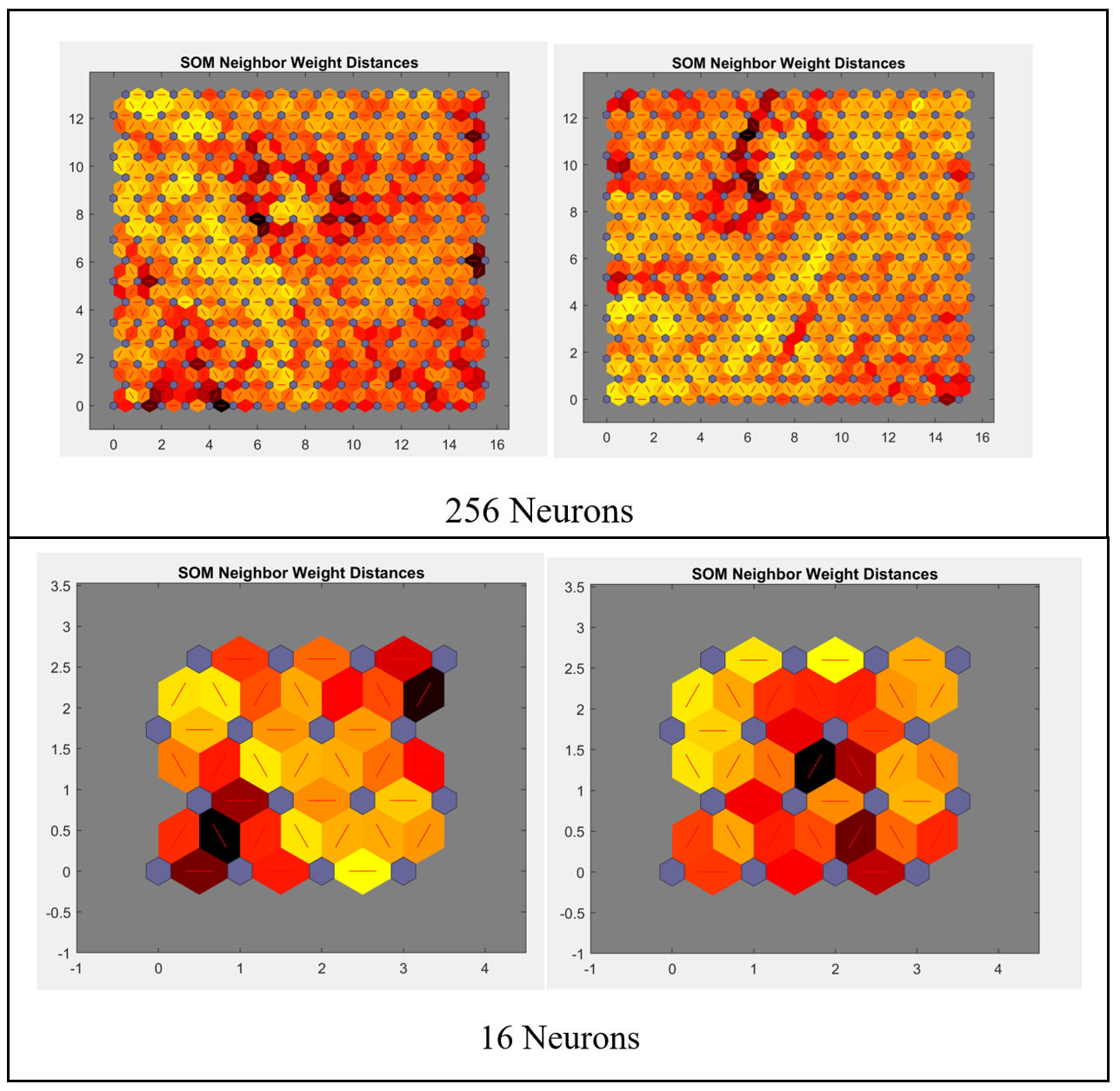}
  \vspace{-3mm}
  \caption{Neighbor Weight Distances}~\label{fig:figure3}
  \vspace{-3mm}
\end{figure}

\begin{figure}[!htb]
    \includegraphics[width=0.9\columnwidth]{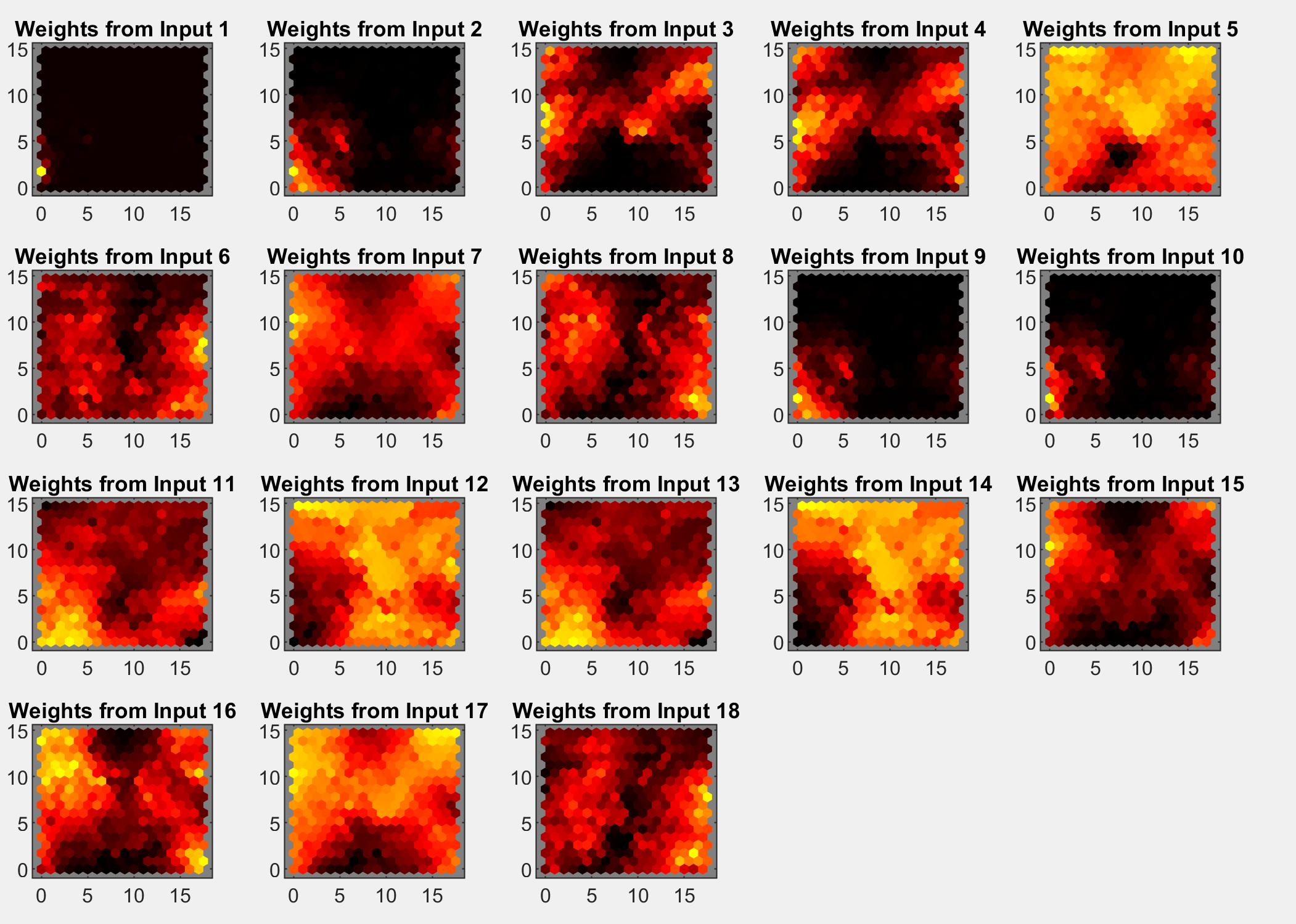}
    \vspace{-3mm}
  \caption{Weight Planes for all Inputs}~\label{fig:figure1}
  \vspace{-3mm}
\end{figure}

\begin{figure}[!htb]
  \includegraphics[width=0.9\columnwidth]{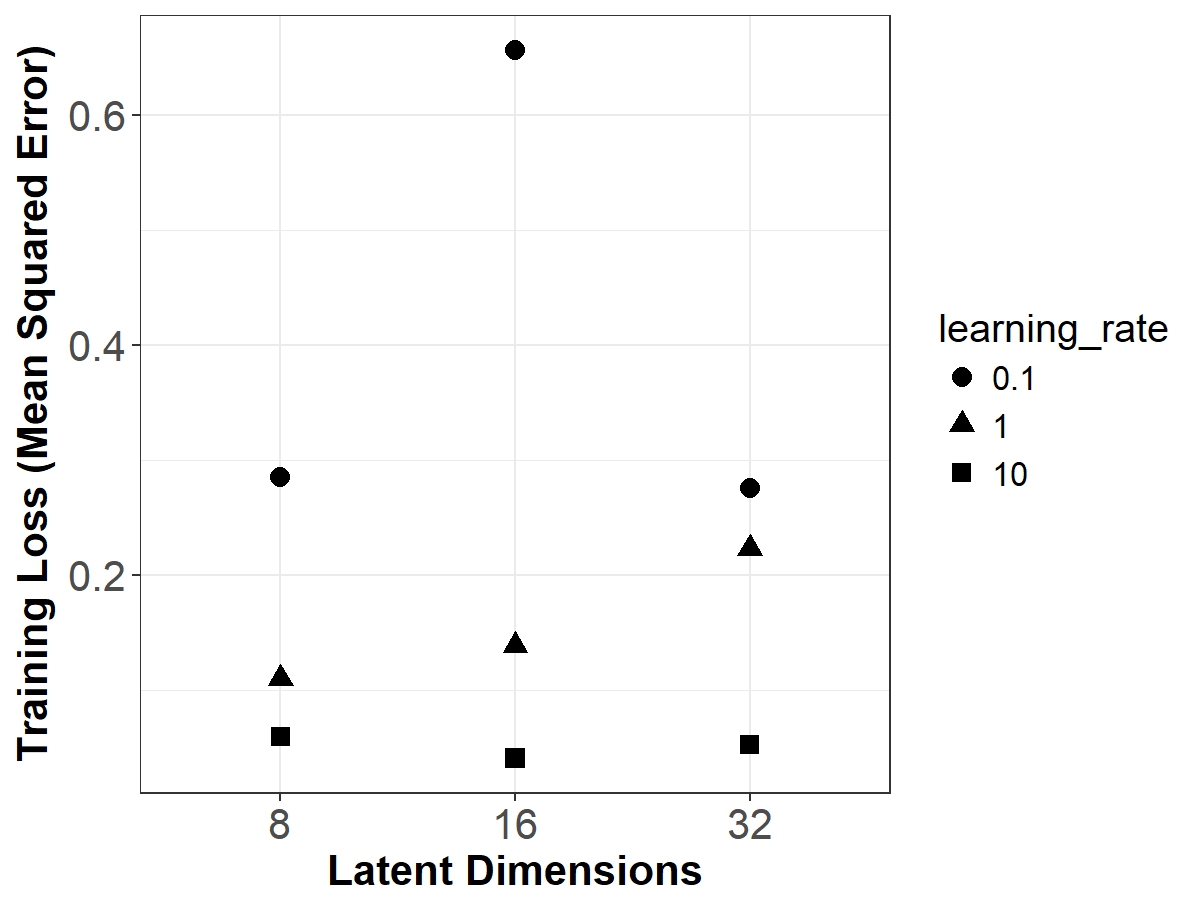}
  \caption{Results of 9 autoencoders to explore parameters selection of the number of latent dimensions and optimizer learning rate}~\label{fig:auto_params}
\end{figure}

\begin{figure}[!htb]
  \includegraphics[width=0.9\columnwidth]{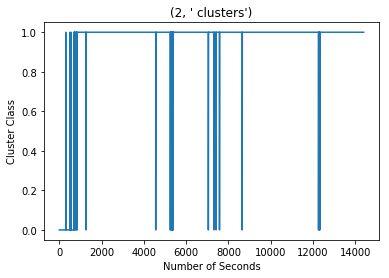}
  \vspace{-3mm}
  \caption{Step plot showing the KMeans model results for 2 clusters. Class 1 is silence and Class 0 is non-silence.}~\label{fig:2clusters}
  \vspace{-3mm}
\end{figure}

\begin{figure}[!htb]
  \includegraphics[width=0.9\columnwidth]{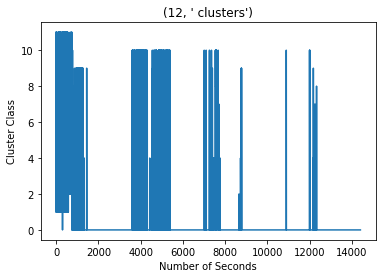}
  \vspace{-3mm}
  \caption{Step plot showing the KMeans model results for 12 clusters. Class 0 is silence and all the other classes are non-silence.}~\label{fig:12clusters}
  \vspace{-3mm}
\end{figure}
 \end{appendices}
\end{document}